\documentstyle[12pt]{article}
\def\beq{\begin{equation}}
\def\eeq{\end{equation}}

\begin{document}

\newcommand{\0}{\mbox{$|0\rangle$}}
\newcommand{\1}{\mbox{$|1\rangle$}}
\newcommand{\2}{\otimes}
\newcommand{\zz}{\mbox{$|z\rangle$}}
\newcommand{\ao}{\mbox{$|a=0\rangle$}}
\newcommand{\ket}{\rangle}
\newcommand{\bra}{\langle}

\begin{center}
{\Large {\bf Quantum Disentanglement and Computation}}\\[8mm]

Asher Peres$^*$ \\[5mm]
{\sl Department of Physics, Technion---Israel Institute of Technology,
32\,000 Haifa, Israel}\\[8mm]

{\bf Abstract}\end{center}

\noindent Entanglement is essential for quantum computation. However,
disentanglement is also necessary. It can be achieved without the need
of classical operations (measurements). Two examples are analyzed: the
discrete Fourier transform and error correcting codes.\\[6mm]

\begin{center}{\bf 1. Introduction}\end{center}

Quantum systems that have interacted are usually described by entangled
wave functions: none of their components has a definite quantum state,
even if the entire composite system is in a pure state. This fact has
been known since the early days of quantum mechanics. For example, the
first reliable calculation of the ground state of the hydrogen molecule
used an entangled wave function~[1]. It was noticed long ago that
entangled quantum systems appear to have paradoxical properties when
their subsystems are measured by independent observers~[2]. Quite
recently, quantum entanglement also found practical applications. These
include secure communication methods, such as quantum cryptography~[3]
and quantum teleportation~[4], and especially quantum computation where
entanglement is pervasive both in computing algorithms~[5] and in error
correcting codes~[6].

It is often stated that after quantum systems interact and become
entangled, it is so difficult to disentangle them that entanglement is
effectively irreversible. In particular, entanglement with an unknown
environment leads to decoherence~[7] and loss of information. In this
article, I shall show that disentanglement is not only possible, but it
is necessary for some aspects of quantum computation. Moreover, contrary
to common wisdom, it can be achieved without measuring the quantum
computer or parts thereof at intermediate stages of the computation.
Two examples are discussed below: the discrete Fourier transform and
error recovery.

The notations used in this article are as follows: the quantum state of
a single logical qubit (quantum binary digit) will be denoted as

\beq \psi=\alpha\,\0+\beta\,\1, \label{qubit}\eeq
where the coefficients $\alpha$ and $\beta$ are complex numbers which
satisfy $|\alpha|^2+|\beta|^2=1$. The symbols \0\ and \1\ represent any
two orthogonal quantum states, such as ``up'' and ``down''  for a spin,
or the ground state and an excited state of a trapped ion. Dirac's ket
notation will in general {\it not\/} be used for generic state vectors
(such as $\psi$) and the $\2$ sign will sometimes be omitted, when the
meaning is clear. Kets will still be used for denoting basis vectors
such as \0\ and \1\ and their direct products.  The latter will be
labelled by binary numbers, such as

\beq |9\ket\equiv|01001\ket\equiv\0\2\1\2\0\2\0\2\1. 
\eeq
Other notations will be explained as needed in the text.\\[5mm]

\begin{center}{\bf 2. Discrete Fourier transform}\end{center}

Our first example of disentanglement is the detection of a periodicity
by means of a discrete Fourier transform. All the numbers that appear
in this section are non-negative integers, with the exception of $\pi$
and $i$.

Let a {\it known\/} function $f(x)$ satisfy

\beq f(x+p)=f(x)\quad{\rm mod} \ N, \eeq
where $N$ is known, but $p$ is not. The problem is to find the period
$p$. For example, in Shor's factoring algorithm~[5], $f(x)=b^x\quad{\rm
mod} \ N,$ where $b$ is given and coprime to $N$ (typically, $N$ and $p$
are huge numbers). We assume that an efficient method is known for
implementing the unitary transformation

\beq |x\ket\,\0\to|x\ket\,|f(x)\ket,\qquad\forall x.\eeq

The quantum computer has two registers. The first one, with $k$ qubits,
may represent any $x$ from 0 to $K-1$ (where $K=2^k\ge2N^2$), and the
second one may represent any $f(x)<N$. The computer is initially
prepared in the state

\beq \psi={1\over\sqrt{K}} \,\sum_{x=0}^{K-1}|x\ket\,|f(x)\ket.\eeq
That state is highly entangled. Note that $f(x)$ takes only $p$
different values, in a periodic way, and that $p\ll K$. It is therefore
convenient to write $\psi$ as

\beq \psi={1\over\sqrt{K}} \,\sum_{c=0}^{p-1} \,\sum_{n=0}^{L(c)}
  |c+np\ket\,|f(c)\ket, \label{psi}\eeq
where $L(c)$ is the largest integer such that $c+Lp<K$.

At this point, the usual way to proceed is to measure the value of the
second register (that is, to test whether each one of its qubits is 0 or
1). This effectively selects one of the $p$ values of $c$, and projects
$\psi$ into a new state

\beq \psi_c={1\over\sqrt{L(c)+1}} \,\sum_{n=0}^{L(c)}
  |c+np\ket\,|f(c)\ket. \eeq
This state is a direct product: the two registers are disentangled, and
as from now the second one can be ignored.

The first register then undergoes a unitary transformation

\beq |x\ket\to U\,|x\ket={1\over\sqrt{K}} \,\sum_{r=0}^{K-1}
  e^{2\pi ixr/K}\,|r\ket, \label{DFT}\eeq
so that $\psi_c$ becomes

\beq \psi'_c=U\,\psi_c={1\over\sqrt{K[L(c)+1]}} \,\sum_{r=0}^{K-1}
\,\sum_{n=0}^{L(c)} e^{2\pi i(c+np)r/K}\,|r\ket.\eeq
The sum over $n$ is a geometric series which can be evaluated
explicitly:

\beq \sum_{n=0}^L \Bigl(e^{2\pi ipr/K}\Bigr)^n=e^{\pi iprL/K}\,
 \,{\sin[\pi pr(L+1)/K]\over\sin(\pi pr/K)}. \label{geom}\eeq

We now measure the first register, which stores a superposition of the
basis vectors $|r\ket$. The probability of getting a particular value of
$r$ is proportional to the square of the fraction on the right hand side
of (\theequation). That probability is sharply peaked when $r$ is near
an integral multiple of $K/p$ (if $K/p$ is an integer, which is then
equal to $L+1$, only exact multiples of $K/p$ are allowed). By repeating
this process sufficiently many times, it is possible to determine $p$
unambiguously~[8].

Note that the disentanglement $\psi\to\psi_c$ was an essentiel step in
the above process. However, achieving that disentanglement does not
require an actual measurement of the second register for selecting a
value of $c$. The same result can be obtained much more easily by {\it
ignoring\/} the second register. The entangled pure state $\psi$ is then
replaced by a reduced density matrix~[9]

\beq \rho={1\over K} \,\sum_{c=0}^{p-1} \,\sum_{n=0}^{L(c)}
\,\sum_{m=0}^{L(c)} |c+np\ket\,\bra c+mp|. \eeq
After we perform the unitary transformation~(\ref{DFT}), namely $\rho\to
U\rho  U^\dagger$, we only have to measure the first register, that is,
to test whether each qubit is up or down. The results of these tests
indicate the value of $r$.

This can also be seen directly from the state $\psi$, without the
density matrix formalism: from (\ref{psi}) and (\ref{DFT}) we have

\beq \psi'={1\over K} \,\sum_{r=0}^{K-1} \,\sum_{c=0}^{p-1}
\,\sum_{n=0}^{L(c)} e^{2\pi i(c+np)r/K}\,|r\ket\,|f(c)\ket. \eeq
The sum over $n$ is again given by (\ref{geom}), and we only have to
measure the first register, while the second one can be ignored.\\[5mm]

\begin{center}{\bf 3. Error correcting codes}\end{center}

Error control is an essential feature of any quantum communication or
computing system. This goal is much more difficult to achieve than
classical error correction, because qubits cannot be read, or copied, or
duplicated, without altering their quantum state in an unpredictable way
[10]. The feasibility of quantum error correction, which for some time
had been in doubt, was first demonstrated by Shor [11]. As in the
classical case, redundancy is an essential element, but this cannot be a
simple repetitive redundancy, where each bit has several identical
replicas and a majority vote is taken to establish the truth.  This is
because qubits, contrary to ordinary classical bits, are usually
entangled.

All quantum error correction methods [6] use several physical qubits for
representing a smaller number of logical qubits (usually a single one).
These physical qubits are prepared in a carefully chosen, highly
entangled state. None of these qubits, taken alone, carries any
information. However, a large enough subset of them may contain a
sufficient amount of quantum information, encoded in relative phases,
for restoring the state of the logical qubits, including their
entanglement with the other logical qubits in the quantum computer. I
shall now review the quantum mechanical principles that make error
correction possible. For simplicity, I shall only consider the simplest
codewords, which represent a single logical qubit.

The state of the entire computer can be written as

\beq \psi=|\alpha\ket\otimes\0+|\beta\ket\otimes\1, \label{computer}\eeq
where one particular qubit has been singled out for the discussion and
appears at the end of each term in $\psi$. The symbols $|\alpha\ket$ and
$|\beta\ket$ represent the collective states of all the other qubits,
that are correlated with \0\ and \1, respectively. In order to encode
the last qubit in Eq.~(\theequation), we introduce an auxiliary system,
called {\it ancilla\/} (this is the Latin word for housemaid). The
ancilla is made of $n$ qubits, initially in a state $|000\ldots\,\ket$.
We shall use $2^n$ mutually orthogonal vectors $|a\ket$, with $a=0$,~1,
\ldots\ (written in binary notation), as a basis for the quantum states
of the ancilla. The labels $a$ are called {\it syndromes\/}, because, as
we shall see, the presence of an ancilla with $a\neq0$ may serve to
identify an error in the encoded system that represents $\psi$.

Encoding is a unitary transformation, $E$, performed on a physical qubit
and its ancilla together:

\beq \zz\otimes\ao\to E\,\Bigl(\zz\otimes\ao\Bigr)\equiv|Z_0\ket,\eeq
where $z$ and $Z$ are either 0 or 1 (the index 0 in $|Z_0\ket$ means
that there is no error at this stage). The unitary transformation $E$
is executed by a quantum circuit (an array of quantum gates). However,
from the theorist's point of view, it is also convenient to consider
$\zz\otimes\ao$ and $|Z_0\ket$ as two different representations of the
same qubit \zz, namely its logical representation and its physical
representation. The first one is convenient for discussing matters of
principle, such as quantum algorithms, while the physical representation
shows how qubits are actually materialized by distinct physical systems,
which may be subject to {\it independent\/} errors. These two different
representations are analogous to the use of normal modes vs.\ local
coordinates for describing the small oscillations of a mechanical
system. One description is mathematically simple, the other one refers
to directly accessible quantities.

Since there are $2^n$ syndromes (including the null syndrome for no
error), it is possible to identify and correct up to $2^n-1$ different
errors that affect the physical qubits, with the help of a suitable
decoding method, as explained below. Let $|Z_a\ket$, with $a=0$,
\ldots\,, $2^n-1$, be a complete set of orthonormal vectors describing
the physical qubits of which the codewords are made: $|0_0\ket$ and
$|1_0\ket$ are the two error free states that represent \0\ and \1, and
all the other $|0_a\ket$ and $|1_a\ket$ are the results of errors
(affecting one physical qubit in the codeword, or several ones, this
does not matter at this stage).  These $|Z_a\ket$ are defined in such a
way that $|0_a\ket$ and $|1_a\ket$ result from definite errors in the
{\it same\/} physical qubits of $|0_0\ket$ and $|1_0\ket$:  for example,
the third qubit is flipped,
${\alpha\choose\beta}\to{\beta\choose\alpha}$, and the seventh one has a
phase error, ${\alpha\choose\beta}\to{\alpha\choose -\beta}$. 

We thus have two complete orthonormal bases, $\zz\otimes|a\ket$ and 
$|Z_a\ket$.  These two bases uniquely define a unitary transformation
$E$, such that

\beq E\,(\zz\otimes|a\ket)=|Z_a\ket, \eeq
and

\beq E^\dagger\,|Z_a\ket=\zz\otimes|a\ket, \label{decod} \eeq
where $a$ runs from 0 to $2^n-1$. Here, $E$ is the encoding matrix, and
$E^\dagger$ is the decoding matrix. If the original and corrupted
codewords are chosen in such a way that $E$ is a real orthogonal matrix
(not a complex unitary one), then $E^\dagger$ is the transposed matrix,
and therefore $E$ and $E^\dagger$ are implemented by the same quantum
circuit, executed in two opposite directions. (If $E$ is complex, the
encoding and decoding circuits must also have opposite phase shifts.)

The $2^n-1$ standard errors $|Z_0\ket\to|Z_a\ket$ are not the only
ones that can be corrected by the $E^\dagger$ decoding. Any error of
type

\beq |Z_0\ket\to U\,|Z_0\ket=\,\sum_a c_a\,|Z_a\ket, \eeq
is also corrected, since

\beq E^\dagger\,\sum_a c_a\,|Z_a\ket=\zz\otimes\,\sum_a c_a\,|a\ket,
\eeq
is a direct product of \zz\ with the ancilla in some irrelevant
corrupted state. Note that no knowledge of the syndrome is needed in
order to correct such errors [12]. Error correction is a logical
operation that can be performed automatically, without having to enter
into the classical world in order to perform quantum measurements. We
definitely know that the error is corrected, even if we don't know the
nature of that error.

It is essential that the result on the right hand side of (\theequation)
be a direct product. Only if the new ancilla state is the same for
$\zz=\0$ and $\zz=\1$, and therefore also for the complete computer
state in Eq.~(\ref{computer}), is it possible to {\it coherently\/}
detach the ancilla from the rest of the computer, and replace it by a
fresh ancilla. (The ancilla may also be restored it to its original
state \ao\ by a dissipative process involving still another, extraneous,
physical system. There is some irony in this introduction of a
dissipative process for stabilizing a quantum computer. The latter was
originally conceived as an analog device with a continuous evolution,
and it is now brought one step closer to a conventional digital
computer!)

This encoding-decoding procedure corrects not only errors in coherent
superpositions of corrupted states, as in (\theequation), but also in
incoherent mixtures. Indeed, if

\beq \rho=\,\sum_j p_j\,\sum_{ab}c_{ja}\,|Z_a\ket\,\bra Z_b|\,c^*_{jb},
\eeq
with $p_j>0$ and $\,\sum p_j=1$, then

\beq E^\dagger\rho\,E= \zz\,\bra z|\otimes
  \,\sum_j p_j\,\sum_{ab}c_{ja}\,|a\ket\,\bra b|\,c^*_{jb}, \eeq
again is a direct product of the logical qubit and the corrupted
ancilla in a mixed state.

In particular, these mixtures include the case where a physical qubit
in the codeword gets entangled with an unknown environment, which is
the typical source of error. Let $\eta$ be the initial, unknown state
of the environment, and let its interaction with a physical qubit
generate the following unitary evolution:

\beq \begin{array}{lll}
 \0\otimes\eta & \to & \0\otimes\mu+\1\otimes\nu,\smallskip\\
 \1\otimes\eta & \to & \0\otimes\sigma+\1\otimes\tau,\end{array}
 \label{error} \eeq
where the new environment states $\mu,\ \nu,\ \sigma$, and $\tau$, are
also unknown, except for unitarity constraints. Now assume that the
physical qubit, which has become entangled with the environment in such
a way, was originally part of a codeword,

\beq |Z_0\ket=|X_{Z0}\ket\otimes\0+|X_{Z1}\ket\otimes\1, \eeq
where the index $Z$ means 0 or 1. (The same index 0 may also refer to
the error free state of a codeword. The interpretation of a subscript 0
should be obvious from the context.) The codeword $|Z_0\ket$, together
with its environment, thus evolves as

\beq Z_0\otimes\eta\to Z'=X_{Z0}\2\Bigl(\0\2\mu+\1\2\nu\Bigr)
  +X_{Z1}\2\Bigl(\0\2\sigma+\1\2\tau\Bigr), \eeq
where I have omitted most of the ket signs, for
brevity. This can also be written as\vspace{-2mm}

\beq \begin{array}{lll}
 Z' & = & 
  \Bigl[X_{Z0}\2\0+X_{Z1}\2\1\Bigr]\,{\displaystyle{\mu+\tau\over2}}+
  \Bigl[X_{Z0}\2\0-X_{Z1}\2\1\Bigr]\,{\displaystyle{\mu-\tau\over2}}\;+
  \smallskip \\ & &
  \Bigl[X_{Z0}\2\1+X_{Z1}\2\0\Bigr]\,{\displaystyle{\nu+\sigma\over2}}+
  \Bigl[X_{Z0}\2\1-X_{Z1}\2\0\Bigr]\,{\displaystyle{\nu-\sigma\over2}}\,.
  \end{array}\eeq
On the right hand side, the vectors

\beq \begin{array}{lll}
 Z_0 & := & X_{Z0}\2\0+X_{Z1}\2\1,\smallskip \\
 Z_r & := & X_{Z0}\2\0-X_{Z1}\2\1,\smallskip \\
 Z_s & := & X_{Z0}\2\1+X_{Z1}\2\0,\smallskip \\
 Z_t & := & X_{Z0}\2\1-X_{Z1}\2\0,\end{array} \label{Z}\eeq
correspond, respectively, to a correct codeword, to a phase error
($\1\to-\1$), a bit error ($\0\leftrightarrow\1$), which is the only
classical type of error, and to a combined phase and bit error.  If
these three types of errors can be corrected, we can also correct any
type of entanglement with the environment, as we shall soon see.

For this to be possible, it is sufficient that the eight vectors in
Eq.~(\ref{Z}) be mutually orthogonal (recall that $Z$ means 0 or 1). The
simplest way of achieving this orthogonality is to construct the
codewords $|0_0\ket$ and $|1_0\ket$ so as to obtain the following scalar
products:

\beq \bra X_{Zy}\,,\,X_{Z'y'}\ket=
 \mbox{$1\over2$}\,\delta_{ZZ'}\,\delta_{yy'}. \label{XX}\eeq
(There are 10 such scalar products, since each index in this equation
may take the values 0 and 1.) If these conditions are satified, the
decoding of $Z'$ by $E^\dagger$ gives, by virtue of Eq.~(\ref{decod}),

\beq E^\dagger\,Z'=\zz\otimes\left(\ao\2{\mu+\tau\over2}
 +|r\ket\2{\mu-\tau\over2}+|s\ket\2{\nu+\sigma\over2}
 +|t\ket\2{\nu-\sigma\over2}\right),\eeq
where $|r\ket$, $|s\ket$, and $|t\ket$ are various corrupted states of
the ancilla. The expression in parentheses is an entangled state of
the ancilla and the unknown environment. We cannot know it explicitly,
but this is not necessary: it is sufficient to know that it is the same
state for $\zz=\0$ and $\zz=\1$, and any linear combination thereof as
in Eq.~(\ref{qubit}). We merely have to discard the old ancilla. It has
been disentangled from the logical qubit.\\[5mm]

\begin{center}{\bf 4. To know or not to know}\end{center}

We have seen that no knowledge of the syndrome is needed in order to
correct an error~[12]. Error correction is a logical operation. It is
part of the software, and can be performed automatically, without
involving irreversible quantum measurements. We can be sure that the
error is corrected, even if we don't know the nature of that error.

The situation is reminiscent of the teleportation of an {\it unknown\/}
quantum state~[4]: the classical information sent by the emitter is not
correlated to the quantum state that she teleports. The receiver too
does not know and cannot know which state he receives, but he can be
sure that this state is identical to the one that was in the emitter's
hands before the teleportation process began. Alternatively,
teleportation can be achieved by standard unitary transformations, for
example by a quantum circuit, without performing any
measurement~[13--15].  Such a circuit can nonetheless be interrupted by
a quantum measurement, and classical information transferred in a
conventional way to another point where the circuit restarts. There, the
classical information is used for performing a unitary transformation
that brings the teleportation process to successful completion.

Likewise, there is no fundamental difference between a ``conscious''
error correction (where the syndrome is actually measured) and an
``unconscious'' one. Conceptually, auto\-matic correction is simpler.
However, it is not obvious that it will also turn out to be simpler from
the point of view of the experimenter. The reason is that a measurement,
namely the conversion of quantum information into classical information,
always starts by a correlation of two quantum systems. In the present
case, when the measured system is a qubit, the measuring apparatus has
two sets of macroscopically distinguishable states, which collectively
behave as the two states of another qubit. The correlation between the
two systems is generated by the familiar controlled-{\sc not} unitary
transformation~[16]. The difference is that in the case of a {\it
measurement\/}, there is no need to maintain the phase coherence of the
two sets of states of the ``target bit'' (the measuring instrument).
Therefore, although conceptually more intricate, this combination of
classical and quantum physics may be easier to realize than a pure
quantum computing system. Similar considerations also apply to the
discrete Fourier transform~[17].\\[5mm]

\noindent{\sl Acknowledgment}. It is a pleasure to dedicate this article
to Rolf Landauer, on the occasion of his 70th birthday. This work was
supported by the Gerard Swope Fund, and the Fund for Encouragement of
Research.\\[5mm]

\begin{center}{\bf References}\end{center}\frenchspacing

\begin{enumerate}
\item N. Rosen, Phys. Rev. {\bf 38}, 2099 (1931).
\item A. Einstein, B. Podolsky, and N. Rosen, Phys. Rev. {\bf 47}, 777
(1935).
\item A. K. Ekert, Phys. Rev. Lett. {\bf 67}, 661 (1991).
\item C. H. Bennett, G. Brassard, C. Cr\'epeau, R.~Jozsa, A.~Peres, and
W.~K.~Wootters, Phys. Rev. Lett. {\bf 70}, 1895 (1993).
\item P. Shor, in Proceedings of the 35th Annual Symposium on the
Foundations of Computer Science, ed. by S. Goldwasser (IEEE Computer
Soc., Los Alamitos, CA, 1994) p.~124.
\item E. Knill and R. Laflamme, Phys. Rev. A {\bf 55}, 900 (1997), and
references therein.
\item W. H. Zurek, Physics Today {\bf 44}, 36 (Oct. 1991).
\item A. Ekert and R. Jozsa, Rev. Mod. Phys. {\bf 68}, 733 (1996).
\item A. Peres, Quantum Theory: Concepts and Methods
(Kluwer, Dordrecht, 1993) p.~122.
\item W. K. Wootters and W. H. Zurek, Nature {\bf 299}, 802 (1982).
\item P. W. Shor, Phys. Rev. A {\bf 52}, 2493 (1995).
\item A. Peres, Phys. Rev. A {\bf 32}, 3266 (1985).
\item S. L. Braunstein, Phys. Rev. A {\bf 53}, 1900 (1996).
\item G. Brassard, in Fourth Workshop on Physics and Computation, ed.
by T. Toffoli, M. Biafore and J. Le\~ao (New England Complex Systems
Institute, 1996), p.~48.
\item M. A. Nielsen and C. M. Caves, Phys. Rev. A {\bf 55}, 2547 (1997).
\item A. Peres, Am. J. Phys. {\bf 42}, 886 (1974); {\bf 54}, 688 (1986).
\item R. B. Griffiths and C.-S. Niu, Phys. Rev. Lett. {\bf 76}, 3228
(1996).
\end{enumerate}
\end{document}